\documentclass[twocolumn,prl,showpacs,preprintnumbers,
               superscriptaddress,floatfix]{revtex4}

\usepackage{amssymb}
\usepackage{graphicx}
\usepackage{dcolumn}
\usepackage{bm}
\usepackage{longtable}

\newcommand{\dfrac}{\displaystyle\frac}
\newcommand{\tN}{{\text{N}}}
\newcommand{\tA}{{\text{A}}}

\begin{document}

\title{Spin symmetry in the anti-nucleon spectrum}

\author{Shan-Gui Zhou}
\affiliation{School of Physics, Peking University,
             Beijing 100871, China}
\affiliation{Max-Planck-Institut f\"ur Kernphysik,
             69029 Heidelberg, Germany}
\affiliation{Institute of Theoretical Physics, Chinese Academy of Sciences,
             Beijing 100080, China}
\affiliation{Center of Theoretical Nuclear Physics, National Laboratory of
             Heavy Ion Accelerator, Lanzhou 730000, China}
\author{Jie Meng}
\affiliation{School of Physics, Peking University,
             Beijing 100871, China}
\affiliation{Institute of Theoretical Physics, Chinese Academy of Sciences,
             Beijing 100080, China}
\affiliation{Center of Theoretical Nuclear Physics, National Laboratory of
             Heavy Ion Accelerator, Lanzhou 730000, China}
\author{P. Ring}
\affiliation{Physikdepartment, Technische Universit\"at M\"unchen,
             85748 Garching, Germany}

\date{\today}

\begin{abstract}
We discuss spin and pseudo-spin symmetry in the spectrum of single
nucleons and single anti-nucleons in a nucleus. As an example we
use relativistic mean field theory to investigate single
anti-nucleon spectra. We find a very well developed spin symmetry
in single anti-neutron and single anti-proton spectra. The
dominant components of the wave functions of the spin doublet are
almost identical. This spin symmetry in anti-particle spectra and
the pseudo-spin symmetry in particle spectra have the same origin.
However it turns out that the spin symmetry in anti-nucleon
spectra is much better developed than the pseudo-spin symmetry in
normal nuclear single particle spectra.
\end{abstract}

\maketitle

Symmetries in single particle spectra of atomic nuclei have been
discussed extensively in the literature, as the violation of
spin-symmetry by the spin-orbit term and approximate pseudo-spin
symmetry in nuclear single particle spectra: atomic nuclei are
characterized by a very large spin-orbit splitting, i.e. pairs of
single particle states with opposite spin ($j=l\pm \frac{1}{2}$)
have very different energies. This fact allowed
the understanding of magic numbers in nuclei and forms the basis
of nuclear shell structure. More than thirty years ago
\cite{AHS.69,HA.69} pseudo-spin quantum numbers have been
introduced by $\tilde{l}=l\pm 1$ and $\tilde{j}=j$ for $j=l\pm
\frac{1}{2}$ and it has been observed that the splitting between
pseudo-spin doublets in nuclear single particle spectra is by an
order of magnitude smaller than the normal spin-orbit splitting.

After the observation that relativistic mean field models yield
spectra with nearly degenerate pseudo spin-orbit
partners~\cite{BBD.95}, Ginocchio showed clearly that the origin
of pseudo-spin symmetry in nuclei is given by a relativistic
symmetry in the Dirac Hamiltonian (\cite{Gin.97,Gin.99} and
references given therein). He found that pseudo-spin symmetry
becomes exact in the limiting case, where the strong scalar and
vector potentials have the same size but opposite sign. However,
this condition is never fulfilled exactly in real nuclei, because
in this limit the average nuclear potential vanishes and nuclei
are no longer bound. It has been found that the quality of
pseudo-spin symmetry is related to the competition between the
centrifugal barrier and the pseudo-spin orbital potential
\cite{MSY.98}.

In relativistic investigations a Dirac Hamiltonian is used. In its
spectrum one finds single particle levels with positive energies
as well as those with negative energies. The latter are
interpreted as anti-particles under charge conjugation. This has
lead to much efforts to explore configurations with anti-particles
and their interaction with nuclei. The possibility of producing a
new kind of nuclear system by putting one or more anti-baryons
inside ordinary nuclei has recently gained renewed interest
\cite{BMS.02}. For future studies of anti-particles in nuclei it
is therefore of great importance to investigate the symmetries of
such configurations.

In a relativistic description nuclei are characterized by two
strong potentials, an attactive scalar field $-S(\bm{r})$ and a
repulsive vectror field $V(\bm{r})$ in the Dirac equation which
for nucleons (labelled by a subscript ``N'') reads,
\begin{equation}
 \left[ \bm{\alpha}\cdot \bm{p} + V_\tN(\bm{r}) + \beta (M-S_\tN(\bm{r}))
 \right] \psi_\tN(\bm{r}) = \epsilon_\tN \psi_\tN(\bm{r}),
 \label{eq:Dirac0}
\end{equation}
where $V_\tN(\bm{r}) = V(\bm{r})$ and $S_\tN(\bm{r}) = S(\bm{r})$.
For a spherical system, the Dirac spinor
$\psi_\tN$ has the form
\begin{equation}
 \psi_\tN (\bm{r},s) = \frac{1}{r}
  \left(
   \begin{array}{c}
    i G_{n\kappa}(r)         Y_{jm}^{l}(\theta,\phi,s )        \\
    - F_{\tilde{n}\kappa}(r) Y_{jm}^{\tilde{l}}(\theta,\phi,s )
   \end{array}
  \right) ,
  \ \ j=l\pm \frac{1}{2},
 \label{eq:SRHspinor}
\end{equation}
where $Y_{jm}^{l}(\theta,\phi)$ are the spin spherical harmonics.
$G_{n\kappa }(r)/r$ and $F_{\tilde{n}\kappa }(r)/r$ form the
radial wave functions for the upper and lower components with $n$
and $\tilde{n}$ radial nodes. $\kappa = \langle 1 +
\mathbf{\bm{\sigma} \cdot l} \rangle = (-1)^{j+l+1/2}(j+1/2)$
characterizes the spin orbit operator and the quantum numbers $l$
and $j$. $\tilde{l}=l-{\text{sign}}(\kappa)$ is the orbital
angular momentum of the lower component. It is therefore well
accepted, that the pseudo-spin quantum number of a particle state
with positive energy are nothing but the quantum numbers of the
lower component~\cite{Gin.97,Gin.99}.

Charge conjugation leaves the scalar potential $S_\tN(\bm{r})$
invariant while it changes the sign of the vector potential
$V_\tN(\bm{r})$. That is, for anti-nucleons (labelled by ``A''),
$V_\tA(\bm{r}) = - V_\tN(\bm{r}) = - V(\bm{r})$ and $S_\tA(\bm{r})
= S_\tN(\bm{r}) = S(\bm{r})$. Charge conjugation of
Eq.~(\ref{eq:SRHspinor}) gives the Dirac spinor for an
anti-nucleon,
\begin{equation}
 \psi_\tA (\bm{r},s) = \frac{1}{r}
  \left(
   \begin{array}{c}
    - F_{\tilde{n}\tilde\kappa}(r) Y_{jm}^{\tilde{l}}(\theta,\phi,s )\\
    i G_{n\tilde\kappa}(r)         Y_{jm}^{l}(\theta,\phi,s )        \\
   \end{array}
  \right) ,
  \ \ j=l\pm \frac{1}{2},
 \label{eq:SRHspinor2}
\end{equation}
with $\tilde\kappa = -\kappa$.

We are only interested in positive energy states of the Dirac
equations. Therefore normal quantum numbers follow the upper
component which is dominant. A particle state is labeled by
$\{nl\kappa m\}$, while its pseudo-quantum numbers are
$\{\tilde{n}\tilde{l} \tilde{\kappa}m\}$. Following
Ref.~\cite{LG.01}, $\tilde{n}=n+1$ for $\kappa >0$; $\tilde{n}=n$
for $\kappa <0$. An anti-particle state is labeled by
$\{\tilde{n}\tilde{l}\tilde{\kappa}m\}$ and its pseudo-quantum
numbers are $\{nl\kappa m\}$. In analogy to Ref.~\cite{LG.01}, we
deduce the relation
\begin{equation}
   n = \tilde{n}+1,\ \text{for}\ \tilde{\kappa}>0;
 \ n = \tilde{n},\ \text{for}\ \tilde{\kappa}<0.
 \label{eq:node2}
\end{equation}
With $\kappa(1-\kappa) = \tilde{l}(\tilde{l}+1)$ and
$\kappa(1+\kappa) = l(l+1)$ in mind, one derives
Schr\"{o}dinger-like equations for the upper and the lower
components
\begin{widetext}
\begin{equation}
 \left[
  - \frac{1}{2M_{+}}
  \left( \frac{d^{2}}{dr^{2}}
        +\frac{1}{2M_{+}}\dfrac{dV_{+}}{dr}\frac{d}{dr}
        -\dfrac{l(l+1)}{r^{2}}
  \right)
  - \dfrac{1}{4M_{+}^{2}} \frac{\kappa}{r} \dfrac{dV_{+}}{dr}
  + M-V_{-}
 \right] G(r)
 = \left\{
    \begin{array}{l}
     + \epsilon_\tN G(r), \\
     - \epsilon_\tA G(r), \\
    \end{array}
 \right.
 \label{eq:origin-of-symmetry1}
\end{equation}
\begin{equation}
 \left[
  - \frac{1}{2M_{-}}
  \left( \frac{d^{2}}{dr^{2}}
        -\frac{1}{2M_{-}}\dfrac{dV_{-}}{dr}\frac{d}{dr}
        +\dfrac{\tilde{l}(\tilde{l}+1)}{r^{2}}
  \right)
  + \dfrac{1}{4M_{-}^{2}}
    \frac{\tilde{\kappa}}{r}\dfrac{dV_{-}}{dr}
  + M-V_{+}
 \right] F(r)
 = \left\{
    \begin{array}{l}
     - \epsilon_\tN F(r), \\
     + \epsilon_\tA F(r), \\
    \end{array}
 \right.
 \label{eq:origin-of-symmetry2}
\end{equation}
\end{widetext}
where $V_{\pm}(r) = V(r)\pm S(r)$ and $M_{\pm} = M_{\pm}(\epsilon)
= M \pm \epsilon \mp V_{\pm}$ with $\epsilon$ = $+\epsilon_\tN$ or
$-\epsilon_\tA$. Both equations are fully equivalent to the exact
Dirac equation with the full spectrum of particle and
anti-particle states. But they carry different quantum numbers.
For particle states the first equation carries spin-quantum
numbers and the second carries pseudo-spin quantum numbers, for
anti-particle states the opposite is true. In the following
discussions we will use either the first or the second equations
according to the type of quantum numbers (spin or pseudo-spin) we
are interested in.

\begin{table}[tbp]
\caption{Relation between symmetry and external fields.}
\label{tab:symmetries}
\begin{ruledtabular}
\begin{tabular}{c|c|c}
             & Particle             & Anti particle        \\
\hline
 $dV_+/dr=0$ & Spin symmetry        & Pseudo spin symmetry \\
\hline
 $dV_-/dr=0$ & Pseudo spin symmetry & Spin symmetry        \\
\end{tabular}
\end{ruledtabular}
\end{table}

\begin{figure}[tbp]
\includegraphics[width=8.0cm]{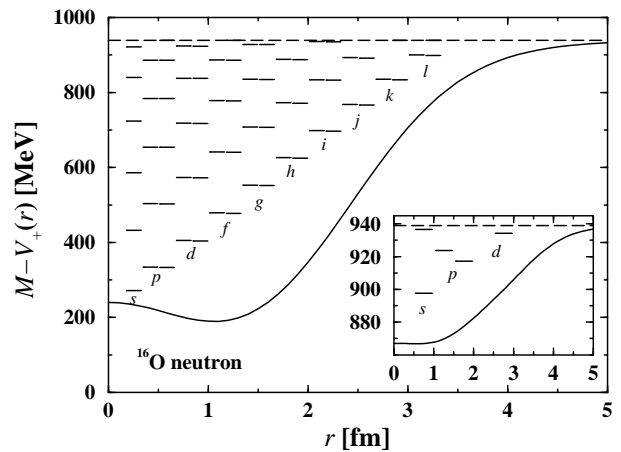}
\caption{Anti-neutron potential and spectrum of $^{16}$O. For each
pair of the spin doublets, the left level is with $\tilde \protect
\kappa < 0$ and the right one with $\tilde \protect \kappa > 0$.
The inset gives neutron potential $M+V_-(r)$ and spectrum.}
\label{fig:spec-o16-neu}
\end{figure}

We give the relation between spin or pseudo-spin symmetry and the
external fields in Table~\ref{tab:symmetries}. If $dV_{+}/dr$ = 0,
we have exact spin symmetry in the particle spectrum and exact
pseudo-spin symmetry in the anti-particle spectrum because states
with the same $l$ (but different $\kappa $) are degenerate in
Eq.~(\ref{eq:origin-of-symmetry1}). $l$ is the orbital angular
momentum of particle states and pseudo orbital angular momentum of
anti-particle states. When $dV_{+}/dr\neq 0$, the symmetries are
broken. But if $dV_{+}/dr$ is so small that the spin-orbit term
(the term $\sim \kappa$) in Eq.~(\ref{eq:origin-of-symmetry1}) is
much smaller than the centrifugal term, there will be approximate
symmetries. For nuclei far from stability where the nuclear
potential is expected to be more diffuse, the spin-orbit splitting
in single nucleon spectra will be also smaller as compared to
stable nuclei. This quenching of the spin-orbit splitting could be
one of the reasons for the change of magic numbers in exotic
nuclei.

Similarly, when $dV_{-}/dr$ = 0 in
Eq.~(\ref{eq:origin-of-symmetry2}), there is an exact pseudo-spin
symmetry in the particle spectra~\cite{Gin.99}. On the other hand,
if we focus on anti-particle states, we have in this case exact
spin symmetry because now $\tilde{l}$ is the orbital angular
momentum. If $dV_{-}/dr \neq 0$ but small, we have approximate
pseudo-spin symmetry in particle spectra and approximate spin
symmetry in anti-particle spectra. This implies that the spin
symmetry in the anti-particle spectrum has the same origin as the
pseudo-spin symmetry in particle spectrum as realized in
Ref.~\cite{Gin.99}. However, there is an essential difference in
the degree to which the symmetry is broken in both cases: the
factor $1/M_{-}^{2} = 1 /(M-\epsilon +V_{-})^{2}$ is much smaller
for anti nucleon states than that for nucleon states. The bound
anti-particle energies $\epsilon_\tA$ are in the region between
$M-V_{+}(0) \lesssim \epsilon_\tA \lesssim M$. For realistic
nuclei roughly we therefore have $0.3$ GeV $ \lesssim \epsilon_\tA
\lesssim 1$ GeV. On the other hand the bound particle states are
in the region of $ M-|V_{-}(0)| \lesssim \epsilon_\tN \lesssim M$,
i.e. for realistic nuclei close to 1 GeV. We therefore have
$|M_{-}(\epsilon_\tA)| > 2|M-S(0)|$ and $|M_{-}(\epsilon_\tN)| <
|V_{-}(0)|$. Thus the factor in front of the $\tilde{\kappa}$-term
is for anti-particle states by more than a factor
$(2|M-S(0)|/|V_{-}(0)|)^{2} \approx 400$ smaller than for particle
states. Spin-symmetry for anti-particle states is therefore much
less broken than pseudo-spin symmetry for particle states.

\begin{figure}[tbp]
\includegraphics[width=8.0cm]{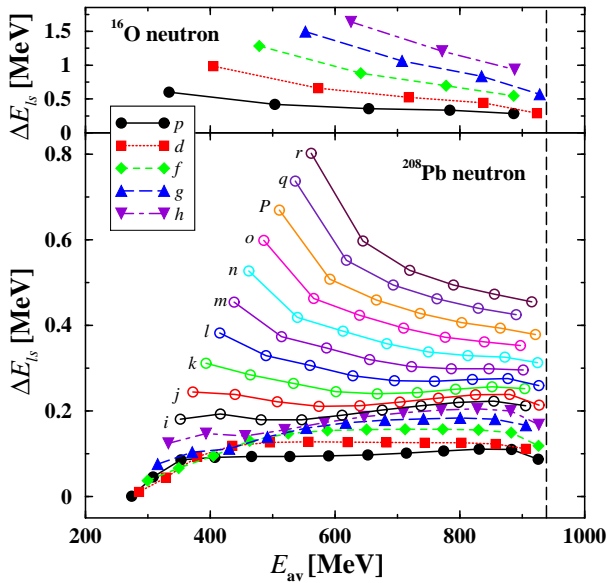}
\caption{Spin-orbit splitting
$\protect\epsilon_\tA(nl_{l-1/2})-\protect\epsilon_\tA(nl_{l+1/2})$
in anti-neutron spectra of $^{16}$O and $^{208}$Pb versus the
average energy of a pair of spin doublets. The vertical dashed
line shows the continuum limit.} \label{fig:spli-neu}
\end{figure}

Since the spin-orbit term in Eq.~(\ref{eq:origin-of-symmetry2}) is
so small for anti-nucleon states, we expect in addition that the
radial wave functions of the spin-doublets are nearly identical,
i.e. the dominant components of spin partners for anti-particle
solutions are much more similar than the small components of
pseudo-spin partners for particles.

Although the present discussion is meant for single particle spectra
in atomic nuclei, the idea is very general. It has been first
discovered that the equality of the vector and scalar potentials
results in spin symmetry in Ref~\cite{ST.71,BR.75} where the
authors suggested applications to meson spectra. However, this
symmetry was only recently found to be valid for mesons with one
heavy quark~\cite{PGG.01}. In the present letter, we illustrate
for the first time in realistic nuclei nearly exact spin symmetry
in the single particle spectra for anti-nucleons. We use for that
purpose non-linear relativistic mean field (RMF) theory
\cite{Rin.96} with modern parameter set NL3. Relativistic Hartree
calculations are carried out in coordinate space for the doubly
magic nuclei $^{16}$O and $^{208}$Pb.

For $^{16}$O, pseudo-spin symmetry cannot be studied successfully
because there are only very few bound nucleon states. However, as
seen in Fig.~\ref{fig:spec-o16-neu}, there are many more
anti-particle states. We find excellent spin symmetry for them.
Since there are too many levels in anti-particle spectra of
$^{208}$Pb (around 400 for either anti-neutrons or anti-protons),
we will not give a similar figure in this case.

In Fig.~\ref{fig:spli-neu} we present the spin-orbit splitting in
anti-neutron spectra of $^{16}$O and $^{208}$Pb. For $^{16}$O, the
spin-orbit splittings are around 0.2-0.5 MeV for $p$ states
($l=1$). With increasing particle number $A$ the spin symmetry in
the anti-particle spectra becomes even more exact. For $^{208}$Pb,
the spin-orbit splittings are $\sim$ 0.1 MeV for $p$ states and
less than 0.2 MeV even for $h$ states ($l=5$) as seen in the lower
panel of Fig.~\ref{fig:spli-neu}. We show in
Table~\ref{tab:ps-splitting-pb208} the pseudo-spin orbit splitting
of the neutron spectrum of $^{208}$Pb to compare them with the
spin-orbit splitting in anti-nucleon spectra. In most cases, the
pseudo-spin orbit splittings for particles are larger than 0.4 MeV
and for deeply bound states, it can reach even values around 4
MeV.

\begin{figure}[tbp]
\includegraphics[width=8.0cm]{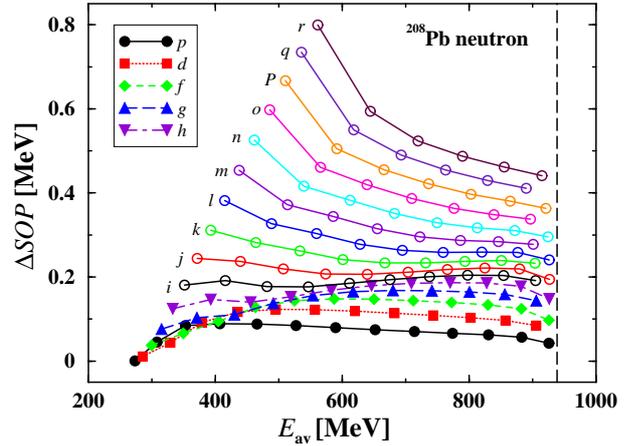}
\caption{Difference of the integration of the spin-orbit potential
$\Delta SOP$ versus the average energy for spin doublets in
$^{208}$Pb. The vertical dashed line shows the continuum limit.}
\label{fig:sop-pb208-neu}
\end{figure}

\begin{table}[tbp]
\caption{Energies (in MeV) of some pseudo-spin doublets in neutron
spectrum of $^{208}$Pb.} \label{tab:ps-splitting-pb208}
\begin{ruledtabular}
\begin{tabular}{ccc|ccc}
 $(n+1)s_{1/2}$ & $nd_{3/2}$ & $\Delta E$ &
 $(n+1)p_{3/2}$ & $nf_{5/2}$ & $\Delta E$ \\
        895.046 &    898.152 &   $-$3.106 &
        904.603 &    908.520 &   $-$3.917 \\
        920.168 &    920.914 &   $-$0.746 &
        929.995 &    930.709 &   $-$0.714 \\
        938.878 &    938.455 &      0.423 &
 $(n+1)f_{7/2}$ & $nh_{9/2}$ & $\Delta E$ \\
 $(n+1)d_{5/2}$ & $ng_{7/2}$ & $\Delta E$ &
        925.638 &    927.984 &   $-$2.346 \\
        914.962 &    918.517 &   $-$3.555 &
 $(n+1)g_{9/2}$ & $ni_{11/2}$& $\Delta E$ \\
        938.484 &    938.292 &      0.192 &
        936.078 &    936.572 &   $-$0.494 \\
\end{tabular}
\end{ruledtabular}
\end{table}

\begin{figure}[tbp]
\includegraphics[width=8.0cm]{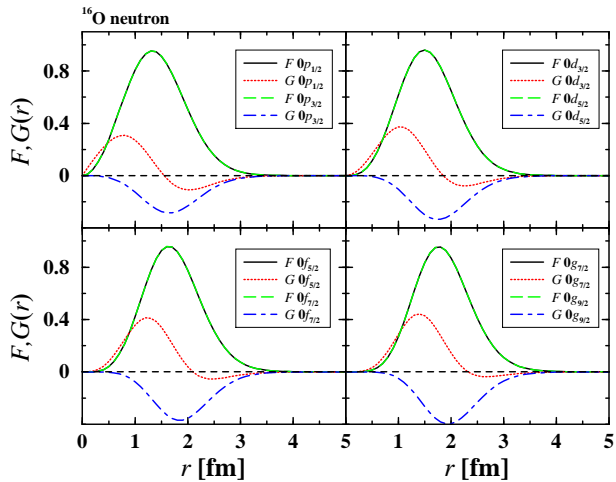}
\caption{Radial wave functions of some spin doublets in the
anti-neutron spectrum of $^{16}$O.}
\label{fig:wf-o16-pdfg}
\end{figure}

\begin{figure}[tbp]
\includegraphics[width=8.3cm]{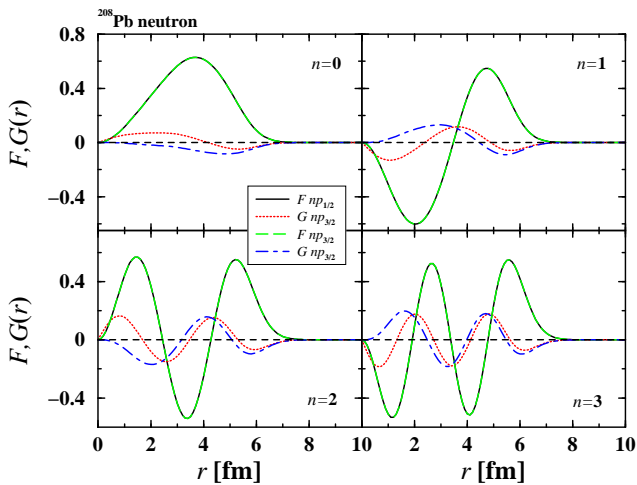}
\caption{Radial wave functions of some spin doublets in the
anti-neutron spectrum of $^{208}$Pb.} \label{fig:wf-pb208-p1234}
\end{figure}

In general, the spin-orbit splitting decreases with the state
approaching the continuum limit. But for very deeply bound
anti-neutron $p$, $d$, $f$ and $g$ states in $^{208}$Pb, the spin
orbit splitting is smaller. This might be due to the competition
between the centrifugal barrier and the spin-orbit potential in
Eq.~(\ref{eq:origin-of-symmetry2}). In order to investigate this
in more detail, we calculated the expectation value of the
spin-orbit potential,
\begin{equation}
 SOP = -\int dr F(r)^{2} \dfrac{1}{4M_{-}^{2}(\epsilon)}
                \dfrac{\tilde{\kappa}}{r} \dfrac{dV_{-}}{dr}.
 \label{eq:sop}
\end{equation}
Since the lower amplitudes of the two spin doublets are nearly
equal to each other (cf. Figs.~\ref{fig:wf-o16-pdfg} and
\ref{fig:wf-pb208-p1234}), we expect the difference, $\Delta SOP$,
gives the main part of $\Delta \epsilon$ of a pair of spin
doublets. In Fig.~\ref{fig:sop-pb208-neu} we present $\Delta SOP$
as a function of the average energy for spin doublets in
$^{208}$Pb. The variational trend of $\Delta SOP$ is roughly in
agreement with that of $\Delta \epsilon$. Particularly, for deeply
bound states, $\Delta SOP\sim \Delta \epsilon $.

Wave functions of pseudo-spin doublets in single nucleon spectra
have been studied extensively in the literature~\cite{Gin.99}. The
lower amplitudes of pseudo-spin doublet are found to be close to
each other. Since the spin symmetry in the anti-nucleon spectrum
is much more exact than the pseudo-spin symmetry in the single
nucleon spectrum, we expect that the upper amplitudes of the spin
doublets coincide with each other even much more.

In Figs.~\ref{fig:wf-o16-pdfg} and \ref{fig:wf-pb208-p1234}, we
show radial wave functions $F(r)$ and $G(r)$ for several
anti-nucleon spin doublets in $^{16}$O and $^{208}$Pb. The
dominant components $F(r)$ are nearly exactly identical for the
two spin partners. On the other hand the small components $G(r)$
of the two spin-partners show dramatic deviations from each other.
The relation between the node numbers of the upper and lower
amplitudes given in Eq.~(\ref{eq:node2}) is seen in
Figs.~\ref{fig:wf-o16-pdfg} and \ref{fig:wf-pb208-p1234}.

In summary, we discussed the relation between the (pseudo)-spin
symmetry in single (anti)-particle states and the external fields
where the (anti)-particle moves. We present the single
anti-nucleon spectra in atomic nuclei as examples and find an
almost exact spin symmetry. The origin of the spin symmetry in
anti-nucleon spectra and the pseudo-spin symmetry in nucleon
spectra have the same origin but the former is much more conserved
in real nuclei. We performed RMF calculations for some doubly
magic nuclei. Even in a very light nucleus, $^{16}$O, the spin
symmetry in the anti nucleon spectrum is very good. The spin
splitting increases with the orbital quantum number and decreases
with the anti-nucleon state approaching the continuum. An
investigation of wave functions shows that the dominant components
of the Dirac spinor of the anti-nucleon spin doublets are almost
identical.

\begin{acknowledgments}
This work was partly supported by the Major State Basic Research
Development Program Under Contract Number G2000077407 and the
National Natural Science Foundation of China under Grant No.
10025522, 10047001 and 10221003 and by the Bundesministerium
f\"{u}r Bildung und Forschung under the project 06 TM 193.
\end{acknowledgments}

\end{document}